\documentclass[a4paper,fleqn,useAMS,usenatbib]{mnras}

\pdfoutput=1

\usepackage{mathptmx}

\usepackage[T1]{fontenc}
\usepackage{ae,aecompl}

\usepackage[dvips]{graphicx}
\usepackage{amsmath}
\usepackage{amsfonts}
\usepackage{amssymb}
\usepackage{comment}
\usepackage[shortlabels]{enumitem}
\usepackage[dvipsnames]{xcolor}
\usepackage[%
        ]{hyperref}
        
\hypersetup{
	colorlinks=true,	
	urlcolor=MidnightBlue,		
	pdfpagelabels=true,
	hypertexnames=true,
	plainpages=false,
	naturalnames=true,
	pdftitle={The SNR of a Transit},    
	pdfauthor={Kipping},     
	linkcolor=WildStrawberry,          
	citecolor=ForestGreen,        
}

\newcommand{\Iout}{ \overline{I_{\mathrm{out}}} }
\newcommand{\Iin}{ \overline{I_{\mathrm{in}}} }

\newcommand{\Tout}{ T_{\mathrm{out}} }
\newcommand{\ti}{ t_{\mathcal{I}} }

\newcommand{\SNR}{ \mathrm{SNR} }

\newcommand{\pdf}{ \mathrm{Pr} }

\title[The SNR of a Transit]{The SNR of a Transit}
\author[Kipping]{David Kipping$^{1}$\thanks{E-mail:
\href{mailto:dkipping@astro.columbia.edu}{dkipping@astro.columbia.edu}}\\
$^{1}$Dept. of Astronomy, Columbia University, 550 W 120th Street, New York NY 10027}

\date{Accepted 2023 May 7. Received 2023 May 2; in original form 2023 April 5}

\pubyear{2023}

\begin{document}
\label{firstpage}
\pagerange{\pageref{firstpage}--\pageref{lastpage}}
\maketitle

\begin{abstract}
Accurate quantification of the signal-to-noise ratio (SNR) of a given
observational phenomenon is central to associated calculations of sensitivity,
yield, completeness and occurrence rate. Within the field of exoplanets, the
SNR of a transit has been widely assumed to be the formula that one would
obtain by assuming a boxcar light curve, yielding an SNR of the form
$(\delta/\sigma_0) \sqrt{D}$. In this work, a general framework is outlined
for calculating the SNR of any analytic function and it is applied to the
specific case of a trapezoidal transit as a demonstration. By refining the
approximation from boxcar to trapezoid, an improved SNR equation is obtained
that takes the form $(\delta/\sigma_0) \sqrt{(T_{14}+2T_{23})/3}$. A solution
is also derived for the case of a trapezoid convolved with a top-hat,
corresponding to observations with finite integration time, where it is proved
that SNR is a monotonically decreasing function of integration time. As a rule
of thumb, integration times exceeding $T_{14}/3$ lead to a 10\% loss in SNR.
This work establishes that the boxcar transit is approximate and it is argued
that efforts to calculate accurate completeness maps or occurrence rate
statistics should either use the refined expression, or even better numerically
solve for the SNR of a more physically complete transit model.
\end{abstract}


\section{Why SNR Matters}
\label{sec:intro}

The expected signal-to-noise ratio (SNR) of a given observational phenomenon is
central to discussions of experimental sensitivity, anticipated yield,
completeness calculations and occurrence rate inferences. One might naively
assume that the SNR of an exoplanetary transit would have a rather trivial
formula, since the transit looks ostensibly similar to a boxcar function.
Indeed, the first transit detection algorithm to gain widespread use, the
Box-fitting Least Squares (BLS) algorithm \citep{kovacs:2002}, explicitly
assumes as such, as well as papers outlining transit surveys, such as
\textit{Kepler} \citep{borucki:2006}. Under this assumption, the SNR of a
single transit takes the form

\begin{align}
\mathrm{SNR} &\propto \delta \sqrt{D},
\end{align}

where $\delta$ is the transit depth and $D$ is the duration of the boxcar
transit, although sometimes the duration is replaced with the number of
points occurring in transit, or something similar (e.g. \citealt{borucki:2006}).
The constant of proportionality here concerns the noise and varies
amongst authors depending on their specific treatment.

In the early years of transit surveys, there was arguably less motivation to
skeptically challenge the boxcar assumption; afterall, the field was in
``gold-rush'' mode and it was not a critical issue if yield estimates were x2
off, or if BLS missed a few planets here and there. But in the
\textit{Kepler}-era, occurrence rate inferences became of central interest; a
calculation which is predicated upon accurate completeness calculations (e.g.
\citealt{petigura:2013,dressing:2013,burke:2015}).

``Completeness'' refers to the true positive rate of a survey, and within the
transit literature is usually framed as the completeness conditional upon the
fact that the planet in question has the correct geometry to transit (which
means the geometric factor is treated separately). Given this framing,
completeness is solely dependent upon the SNR of the signal in question - 
the higher the SNR, the higher the expected completeness.

Some of the earliest occurrence rate calculations assumed that the completeness
was a Heaviside step function with respect to SNR. For example,
\citet{howard:2012} assumed that all transits with SNR$>10$ have 100\%
completeness. \citet{fressin:2013} introduced the more sophisticated
linear-ramp model, where completeness was 0\% up to some lower threshold SNR,
then linearly ramped up to 100\% where it saturated. \citet{christiansen:2013}
advanced this further still, matching injection recovery simulations to a
continuous, smooth inverse gamma cumulative density function. Modern
calculations (e.g. see \citealt{hsu:2019}) often calculate completeness scores
bespoke to each target \citep{burke:2017}. Regardless, at the core of all of
these approaches is an SNR calculation. For example, if we took a given time
series with stationary noise, and injected two different planets but with
the same SNR, they should have the same completeness.

Clearly, an accurate SNR calculation is an essential ingredient, yet this term
has received minimal attention compared to the parameterisation of the
completeness curve. Indeed, even relatively modern \textit{Kepler}
occurrence rate calculations still explicitly adopt a SNR formula that
implicitly assumes a boxcar transit. For example, see Equation~(8) of
\citet{hardegree:2019} and Equation~(4) of \citet{hsu:2019}. The issue
is likely even more widespread as authors don't always state how they
define SNR in similar studies (e.g. \citealt{grunblatt:2019}).

Besides from occurrence rate calculations, the boxcar approximation is
also widely assumed in survey yield calculations. For example,
\citet{beatty:2008} explicitly state that a boxcar model is assumed
(see text below their Equation~4), an assumption also present
in \citet{sullivan:2015} (see second sentence of their Section~6) and
\citet{barclay:2018} (see their Equation~2).

The precise consequences of making this simplifying assumption upon these
yield and occurrence rate calculations is non-trivial and beyond the scope of
this work, but we can be certain that no exoplanet truly produces a boxcar
transit. In this work, we ask then - what is the SNR of a transit?

Section~\ref{sec:chi2} describes a general framework for calculating SNR.
Section~\ref{sec:conventional} (\ref{sec:binning}) applies this to the specific
case of a (convolved) trapezoidal light curve, rather than a limb darkened
\citet{mandel:2002} light curve. Despite this limitation, our work shows
that even this is sufficient to establish that boxcar transit approximation
should not be used and Section~\ref{sec:discussion} expands upon this by
discussing numerical approaches that can be used in real world cases.

\section{A Rigorous Definition for SNR}
\label{sec:chi2}

Chi-squared ($\chi^2$) is a foundational concept in the statistical assessment
of fit quality, applicable for independent normally distributed noise. The
$\chi^2$ metric is defined as the sum of the squares of the normalised
residuals, where the residuals here are the differences between the data
($\mathbf{y}$) and some hypothesised model ($f[\mathbf{t}]$), with
the normalisation being with respect to measurement uncertainties
($\boldsymbol{\sigma}$):

\begin{align}
\chi^2 &= \sum_{j=1}^n \Big(\frac{y_j-f[t_j]}{\sigma_j}\Big)^2.
\end{align}

By construction, it only makes sense to speak of a $\chi^2$ conditional upon
some model. For data of length $n=1$, it's easy to see how this means that
$\sqrt{\chi^2}$ is the absolute difference between the model and the data in
units of sigma (the standard error). And indeed, that statement effectively
holds for any $n$, and the net number of sigmas deviance between the model and
the data is $\sqrt{\chi^2}$.

With this in mind, we can employ a SNR definition that equals the number of
sigmas deviance between a hypothesised model, $\mathbb{M}$, and a null model,
$\mathbb{N}$. We emphasise that this definition treats SNR as an act of model
comparison. But ultimately, all detections are in fact an act of model
comparison, something easy to forget since often we negate to articulate the
tacit null model. To use an analogy, how can one ever measure the height of a
mountain unless one compares it to some contextual plain? In this definition,
our SNR becomes

\begin{align}
\mathrm{SNR}_{\Delta\chi^2} &= \sqrt{ \chi_{\mathbb{N}}^2 - \chi_{\mathbb{M}}^2 }.
\label{eqn:SNRchi2eq}
\end{align}

Whilst the above offers a conceptual outline of the relationship between SNR
and $\chi^2$, a more rigorous justification is provided in \citet{scharf:1991}
and \citet{haykin:1994}. It's worth highlighting that this is hardly a novel
statement, even within the field of exoplanets, for example
\citet{zakamska:2011} define SNR using the same principle concerning
simulated radial velocity observations.

\subsection{From the hypothesized model}

Consider the $\chi^2$ resulting from model $\mathbb{M}$ first, which predicts
the $\mathbf{y}$ data to be described by $f_{\mathbb{M}}[\mathbf{t}]$. In both
this subsection and the next, we will assume that the data is distributed as
$y_j \sim \mathcal{N}(f_{\mathbb{M}}[t_j],\sigma_j)$ i.e. the data is normally
distributed about the hypothesised model. This is an ideal case where the fit
model and the generative model are the same i.e. we are using the correct
model. The presence of any latent unmodeled processes would naturally
attenuate the SNR presented here.

In what follows, we derive the expectation value for the $\chi^2$ arising from
the null model. This is a standard and well-known derivation (e.g. see
\citealt{bain:1992}; \citealt{grimmett:2001}), but we walk through it here
for the pedagogical clarity and to allow us to maintain a consistent notation
system throughout.

Let us define the residuals resulting from the proposed model,
$\mathbb{M}$ as $\mathbf{r}_{\mathbb{M}}$, for which we can now
state that $r_{\mathbb{M},j} \sim  \mathcal{N}(0,\sigma_j)$. If we further
define that the normalized residuals are given by $\rho_{\mathbb{M},j} =
(r_{\mathbb{M},j}/\sigma_j)$, then it follows that $\rho_{\mathbb{M},j} \sim
\mathcal{N}(0,1)$ i.e. a standard normal. We may now write that the square
of the normalized residual is given by $u_{\mathbb{M},j} =
\rho_{\mathbb{M},j}^2$ as an intermediate step towards
$\chi_{\mathbb{M}}^2$($=\sum u_{\mathbb{M},j}$). Because of the square, there
are two roots for $\rho_{\mathbb{M},j}$, specifically
$\rho_{\mathbb{M},j}=\pm\sqrt{u_{\mathbb{M},j}}$, and thus this factor needs to
be accounted for when we transform the $\pdf(\rho_{\mathbb{M},j})$ distribution
to $\pdf(u_{\mathbb{M},j})$, such that

\begin{align}
\pdf(u_{\mathbb{M},j})\,\mathrm{d}u_{\mathbb{M},j} &= 2 \pdf(\rho_{\mathbb{M},j}) \Big| \frac{\mathrm{d}\rho_{\mathbb{M},j}}{\mathrm{d}u_{\mathbb{M},j}}\Big|\,\mathrm{d}u_{\mathbb{M},j}.
\end{align}

Since $\mathrm{d}\rho_{\mathbb{M},j}/\mathrm{d}u_{\mathbb{M},j} =
\tfrac{1}{2}u_{\mathbb{M},j}^{-1/2}$ and $\pdf(\rho_{\mathbb{M},j}) =
\tfrac{1}{\sqrt{2\pi}} \exp(-\tfrac{1}{2}\rho_{\mathbb{M},j}^2)$ (a standard
normal) then this yields

\begin{align}
\pdf(u_{\mathbb{M},j}) &= \frac{1}{\sqrt{2\pi u_{\mathbb{M},j}}} \exp\Big( -\frac{u_{\mathbb{M},j}}{2} \Big),
\end{align}

which is the chi-squared distribution for one degree of freedom. To go from
$u_{\mathbb{M},j} \to \sum u_{\mathbb{M},j}$($=\chi_{\mathbb{M}}^2$) we need
to add the random variates together. This can be achieved by noting that
the distribution describing the sum of variates has a characteristic function
equal to the product of the characteristic functions from the individual
variates being summed. As a first step towards this goal, the characteristic
function of $\pdf(u_{\mathbb{M},j})$ is found through a Fourier transform:

\begin{align}
\phi_{\mathbb{M},j} &= \int_{0}^{\infty} \pdf(u_{\mathbb{M},j}) \exp(i u_{\mathbb{M},j} t) \mathrm{d}u_{\mathbb{M},j},\nonumber\\
\qquad&= \frac{1}{\sqrt{1-2it}}.
\end{align}

Accordingly, the characteristic function of $\pdf(\chi_{\mathbb{M}}^2)$ is
given by

\begin{align}
\Phi_{\mathbb{M}} &= \prod_{j=1}^n \phi_{\mathbb{M},j},\nonumber\\
\qquad&= (1-2it)^{-n/2}.
\end{align}

One could transform back to the probability density function for
$\chi_{\mathbb{M}}^2$ at this point, using an inverse Fourier transform;
however, in this work we really only care about the expectation value of
$\chi_{\mathbb{M}}^2$ ($=\mathbb{E}[\chi_{\mathbb{M}}^2]$). This can be
found by simply calculating the first raw moment of the distribution,
which is given by

\begin{align}
\mathbb{E}[\chi_{\mathbb{M}}^2] &= \frac{\partial \Phi_{\mathbb{M}}}{\partial (it)}\Big|_{t=0}\nonumber\\
\qquad&= n.
\label{eqn:poschi2}
\end{align}

This result slightly differs from the more conventional framing of $\chi^2$
as having a mean at the \# of degrees of freedom (= \# of data points - \# of
free parameters), whereas here $n$ is just the number of data points. That's
because our framing of the problem here does not involve a regression
(hence there's no defined value of \# of free parameters anyway). Instead,
we are calculating the expectation value of the probability distribution for
the normalised residuals, where the residuals are defined as the difference
between the data and \textit{the generative model} (rather than \textit{a
regressed model}). In the limit of large data $(n\gg1)$, which we adopt in this
work, the definitions are of course equivalent and this follows from the fact
that the regression will asymptotically approach the truth in this limit. It's
somewhat advantageous to frame the problem in this way to generalise our work
to non-linear models (such as a transit model, including even the box-car
simplification scheme) for which the number of free parameters is ill-defined
\citep{andrae:2010}, or indeed linear models for which the basis functions are
not linearly independent or have influential priors.

\subsection{From the null model}

We now proceed to calculate the mean $\chi^2$ resulting from the null model
i.e. $\chi_{\mathbb{N}}^2$. Let $r_{\mathbb{N},j} = (y_j -
f_{\mathbb{N}}[t_j])$ be the residuals between the data and the null model.
Recall that $y_j \sim \mathcal{N}(f_{\mathbb{M}}[t_j],\sigma_j)$, which
means that $r_{\mathbb{N},j} \sim
\mathcal{N}(f_{\mathbb{M}}[t_j]-f_{\mathbb{N}}[t_j],\sigma_j)$.

We can make further progress by normalizing by $\sigma_j$ as before, such that
$\rho_{\mathbb{N},j}=r_{\mathbb{N},j}/\sigma_j$, and thus allowing us
to write that $\rho_{\mathbb{N}I,j} \sim \mathcal{N}(\Delta_j,1)$, where
we define $\Delta_j = (f_{\mathbb{M}}[t_j]-f_{\mathbb{N}}[t_j])/\sigma_j$.

With the probability distribution for $\rho_{\mathbb{N}I,j}$ obtained, we can
now transform to that for $u_{\mathbb{N},j} = \rho_{\mathbb{N},j}^2$ similar
to before, but instead of obtaining the chi-squared distribution with one
degree of freedom, this defines the \textit{non-central} chi-squared
distribution with one degree of freedom - and a noncentrality parameter given
by $\Delta_j^2$:

\begin{align}
\pdf(u_{\mathbb{N},j}) &= \frac{1}{\sqrt{2\pi u_{\mathbb{N},j}}} \exp\Big(\frac{-u_{\mathbb{N},j}^2-\Delta_j^2}{2}\Big) \cosh (\sqrt{u_{\mathbb{N},j}}\Delta_j)
\end{align}

The characteristic function of the non-central chi-squared distribution of
one degree of freedom is given by

\begin{align}
\phi_{\mathbb{N},j} &= \frac{1}{\sqrt{1-2it}} \exp\Big(\frac{i t \Delta_j^2}{1-2 i t}\Big).
\end{align}

From this, the characteristic function for $\chi_{\mathbb{N}}^2$ is

\begin{align}
\Phi_{\mathbb{N}} = \sum_{j=1}^{n} \phi_{\mathbb{N},j}.
\end{align}

Differentiating to obtain the first moment as before, which here now equals
$\mathbb{E}[\chi_{\mathbb{N}}^2]$, we obtain

\begin{align}
\mathbb{E}[\chi_{\mathbb{N}}^2] &= n + \sum_{j=1}^{n} \Delta_j^2.
\label{eqn:nullchi2}
\end{align}

\subsection{The SNR for Continuous Functions}
\label{sub:SNRcontinuous}

The SNR can now be calculated by our definition earlier in
Equation~(\ref{eqn:SNRchi2eq}). Specifically, we have

\begin{align}
\mathrm{SNR}_{\Delta\chi^2} &= \sqrt{ \sum_{j=1}^{n} \Delta_j^2 },\nonumber\\
\qquad&= \sqrt{ \sum_{j=1}^{n} \Big(\frac{f_{\mathbb{M}}[t_j] - f_{\mathbb{N}}[t_j]}{\sigma_j}\Big)^2 }.\nonumber\\
\end{align}

Whilst the above is fairly intuitive, it's also somewhat limited in being
dependent upon choices regarding cadence and sampling. The final step is to now
generalise the above to an arbitrary continuous function (although in our case
we care about the transit function specifically). To do so, let us assume that
the sampling is regular with no data gaps, no read time (such that exposure
time equals cadence) and is also homoscedastic. We further assume that
the sampling is sufficiently dense that we can ignore the effect of the phasing
of the data itself with respect to the mid-transit time. Since the uncertainty
on the data will depend on possible binning choices, we define the uncertainty
expected over one unit of time as $\sigma_0$. By Poisson statistics, the
uncertainty over three units of time will this be $\sigma_0/\sqrt{3}$, or more
generally the uncertainty over a time $\delta t$ cadence will be
$\sigma_0/\sqrt{\delta t}$, such that

\begin{align}
\mathrm{SNR}_{\Delta\chi^2} &= \sqrt{ \sum_{j=1}^{n} \Delta_j^2 },\nonumber\\
\qquad&= \sqrt{ \sum_{j=1}^{n} \Big(\frac{f_{\mathbb{M}}[t_j] - f_{\mathbb{N}}[t_j]}{\sigma_0}\Big)^2 \delta t }.
\label{eqn:discrete}
\end{align}

In the limit of infinitesimal $\delta t$, this becomes

\begin{align}
\mathrm{SNR}_{\Delta\chi^2} &= \sqrt{ \int_{t=t_1}^{t_2}  \Big(\frac{f_{\mathbb{M}}[t] - f_{\mathbb{N}}[t]}{\sigma_0}\Big)^2\,\mathrm{d}t },
\label{eqn:SNRchi2}
\end{align}

where $(t_2-t_1)$ is the total temporal length of observations.

\subsection{Modification for Linear Regression Problems}

In the real world, the exact generative model behind the data is unknown to us,
and indeed inaccessible. However, if at least the parametric form of the model
is known, then the regressed model will tend towards the truth asymptotically
as the data accumulates. Nevertheless, let us revisit our earlier assumption
concerning this point. For a linear model, with no priors and a full rank
design matrix, regression of said model to data will not result in
$\mathbb{E}[\chi_{\mathbb{M}}] = n$, but rather
$\mathbb{E}[\chi_{\mathbb{M}}] = n-k$, where $k$ is the rank of the design
matrix \citep{andrae:2010}. We highlight that whether the transit model be
\citet{mandel:2002}, a trapezoid or even a box-car, none of them are linear
models and what follows is not actually applicable to the transit scenario.
Regardless, let us proceed to gain some insights about the consequences of
this earlier assumption.

From Equation~(\ref{eqn:nullchi2}), one can see that nominally the expected
$\chi^2$ value is the $n$ plus the sum of the square deviances between the
model being tested (in that case the null model) and the generative model.
If the model being tested is the generative model, then the deviances are
zero and hence Equation~(\ref{eqn:nullchi2}) becomes
Equation~(\ref{eqn:poschi2}). Accordingly, this means that the expected
chi-squared values get modified, in the case of regression, to

\begin{align}
\chi_{\mathbb{M}}^2 &\to n - k_{\mathbb{M}},\nonumber\\
\chi_{\mathbb{N}}^2 &\to n - k_{\mathbb{N}} + \sum_{j=1}^{n} \Delta_j^2,
\end{align}

where $k_{\mathbb{M}}$ is the rank of the design matrix for model
$f_{\mathbb{M}}[t]$ and $k_{\mathbb{N}}$ is that for model
$f_{\mathbb{N}}[t]$. The modification to our SNR equation is then
simply 

\begin{align}
\mathrm{SNR}_{\Delta\chi^2} &= \sqrt{ (k_{\mathbb{N}}-k_{\mathbb{M}}) + \int_{t=t_1}^{t_2}  \Big(\frac{f_{\mathbb{M}}[t] - f_{\mathbb{N}}[t]}{\sigma_0}\Big)^2\,\mathrm{d}t }.
\end{align}

Since one usually would have $k_{\mathbb{M}} > k_{\mathbb{N}}$, the
new term included is negative and thus serves as a penalty to the SNR.
This again reinforces our claim that the SNR derived in the previous
subsection and presented in Equation~(\ref{eqn:SNRchi2}) is an idealised,
upper-limit on SNR.

\section{The SNR of a Trapezoidal Transit}
\label{sec:conventional}

\subsection{The Double Box (BB) Approximation}
\label{sub:BB}

To zeroth order approximation, the shape of a transit light is a box and the
appeal of this model is undoubtedly borne of its simplicity. Let us call this
the double box (BB) approximation due its double appearance in that statement;
both as the assumed generative model and the model employed to define SNR. The
fact that our generative model is here assumed to be a box immediately makes it
incongruous with this section, whose title clearly seeks the SNR of a
\textit{trapezoidal} transit. Nevertheless, let us continue with this case for
the moment as it provides some context for what follows and helps us understand
the origin of the most commonly cited formulae for transit SNR.

In this scenario, it is straight-forward to calculate the SNR. Let us begin by
first defining the transit model as a function of time from the mid-transit,
$t$, writing the normalised intensity as a function of time as

\begin{equation}
f_{\mathrm{B}}[t] =
\begin{cases}
1  & \text{if } t \leq -D/2,\\
1 - \delta  & \text{if } -D/2 < t \leq D/2,\\
1  & \text{if } t > D/2,
\end{cases}
\label{eqn:fbox}
\end{equation}

where $\delta$ is the depth of the transit and $D$ is the duration of the
box-car. Now it is necessary to define the SNR of this transit. The truth is
that there's no single way to define SNR, and our exploration of the literature
reveals it is rarely rigorously derived or even defined. In the case of the BB
approximation, however, two reasonable proposals yield the same result, which
ultimately stem from the self-consistency of the double box framework. The
first, and perhaps most common, is to define SNR as the depth divided by the
error on the depth. So we now need expressions for each of these components.

Looking at the box-car function, Equation~(\ref{eqn:fbox}), one can
re-arrange to solve for depth using $\delta = (\Iout - \Iin)$, where $\Iout$
is the mean out-of-transit intensity and $\Iin$ is the the mean in-transit
intensity. In the more general case where we observe unnormalised intensities,
this becomes 

\begin{align}
\Delta F_{\mathrm{BB}} \equiv \frac{\Iout - \Iin}{\Iout},
\end{align}

where i) $\delta$ is replaced with ``$\Delta F$'' to define the flux change
(which will be useful later to avoid notational confusion) and ii) the ``BB''
subscript is added to explicate the transit model was generated using a box-car
function, Equation~(\ref{eqn:fbox}), \textit{and} the depth is solved for
assuming a boxcar function by inverting Equation~(\ref{eqn:fbox}).

Under the assumption of no time-correlated noise structure, the uncertainty on
$\Delta F_{\mathrm{BB}}$ will be

\begin{align}
\sigma_{\Delta F_{BB}}^2 &= (\partial_{\Iout} \delta_{BB})^2 \sigma_{\Iout}^2 + (\partial_{\Iin} \delta_{BB})^2 \sigma_{\Iin}^2,\nonumber\\
\qquad&= \frac{ \Iout^2 \sigma_{\Iin}^2 + \Iin^2 \sigma_{\Iout}^2 }{ \Iout^4 }.
\end{align}

To make progress, it is assumed that the noise is dominated by photon-noise and
thus can apply Poisson noise statistics. In this case, if the noise per unit
time is $\sigma_0$, then the noise over an interval of $D$ (the transit
duration) will be $\sigma_{\Iin} = \sigma_0/\sqrt{D}$. Similarly,
$\sigma_{\Iout} = \sigma_0/\sqrt{\Tout}$, where $T_{\mathrm{out}}$ is the
out-of-transit observational baseline. Now applying this to normalised
intensities using Equation~(\ref{eqn:fbox})

\begin{align}
\mathrm{SNR}_{\Delta F,BB} &= \frac{\delta}{\sigma_0} \frac{1}{\sqrt{ \frac{1}{D} + \frac{(1-\delta)^2}{\Tout} }},
\end{align}

and in the limit of large amounts of out-of-transit data

\begin{align}
\lim_{\Tout \to \infty} \mathrm{SNR}_{\Delta F,BB} &= \frac{\delta}{\sigma_0} \sqrt{D}.
\label{eqn:SNRBB}
\end{align}

An alternative definition for SNR is the $\SNR_{\Delta\chi^2}$ formula
presented earlier in Equation~(\ref{eqn:SNRchi2}). Feeding
Equation~(\ref{eqn:fbox}) into Equation~(\ref{eqn:SNRchi2}), we obtain

\begin{align}
\mathrm{SNR}_{\Delta\chi^2} &= \frac{\delta}{\sigma_0} \sqrt{D},
\end{align}

i.e. the same result. This is not surprising given that model is completely
defined by the step function in intensity and we are here comparing the same
model as the generative one. As an aside, it is noted that if one phase-folds
$N$ transits together (or equivalently regresses a multi-epoch light curve
model to $N$ transits), then $\sigma_0 \to \sigma_0/\sqrt{N}$.

Equation~(\ref{eqn:SNRBB}) is indeed the most commonly quoted definition for
SNR of a transit. The obvious major drawback of this formula is that its
underlying assumptions are patently wrong, a box-car transit is impossible
(and has even been suggested as a technosignature given its impossibility;
\citealt{cloaking:2016}). It really doesn't mean much to speak of this SNR,
since it corresponds to a case which never exists - a generative model equal
to a box-car transit. In this sense, it cannot be compared to other SNRs values
presented in the next two subsections, since they are for a distinct generative
model (namely a trapezoid).

It is noted that this formula also appears in \citet{carter:2008}, despite the
fact the authors are considering the SNR of $\delta$ from a trapezoidal
transit, not a boxcar. In that work, the authors refer to the SNR as ``$Q$''
and after notational translation have the same result, except that $D \to W$
which is not a general duration but fixed to the full width half maximum
transit duration i.e.

\begin{align}
W \equiv \frac{T_{14}+T_{23}}{2}.
\end{align}

\subsection{The Trapezoid-Box (TB) Approximation}

Things become more complicated when one considers the case where the assumed
generative model is a trapezoid, but SNR is still in terms of
a box - what it is dubbed here the trapezoid-box (TB) approximation. In this
scenario, the generative model is now that of a trapezoid of full transit
duration $T_{14}$ and totality duration $T_{23}$:

\begin{equation}
f_{\mathrm{T}}[t] =
\begin{cases}
1  & \text{if } t \leq -T_{14}/2,\\
1 - \frac{(2t+T_{14})\delta}{T_{14}-T_{23}} & \text{if } -T_{14}/2 < t \leq -T_{23}/2,\\
1 - \delta & \text{if } -T_{23}/2 < t \leq T_{23}/2,\\
1 + \frac{(2t-T_{14})\delta}{T_{14}-T_{23}} & \text{if } T_{23}/2 < t \leq T_{14}/2,\\
1  & \text{if } t > T_{14}/2.
\end{cases}
\label{eqn:ftrap}
\end{equation}

A basic issue now is that when placing a box over the trapezoid,
$\SNR_{\mathrm{TB}}$ must be sensitive to how big we make that box, i.e. the
value of $D$. Whilst that was also true in the previous section, there was only
really one obvious choice for $D$ to set it to be equal to the duration used in
the generative model i.e. $D$. But now the question arises, should $D =
T_{23}$? Or maybe $T_{14}$? Or perhaps some combination of the two? In this
sense, there is no single solution for $\SNR_{\mathrm{TB}}$.

A case study example of this comes from \citet{ks:2016}. In that work, I
too casually suggested that SNR could be defined using the $\delta$ method
applied to a trapezoid. Further, $D$ was set to $T_{14}$. These were choices
made by myself as the lead author and in hindsight they were made too hastily;
but afterall the process of science is one of continuous refinement, even in
our own understanding. The choice of $D$ sets what defines $\Iout$ and $\Iin$.
Let's proceed first by keeping $D$ general and following the $\Delta F$ method,
such that

\begin{align}
\overline{I_{\mathrm{in},\mathrm{TB}}} &= D^{-1} \int_{-D/2}^{D/2} f_{\mathrm{T}}[t]\,\mathrm{d}t
\end{align}

yielding

\begin{equation}
\overline{I_{\mathrm{in},\mathrm{TB}}} =
\begin{cases}
1-\delta  & \text{if } D \leq T_{23},\\
1+\delta \big(\frac{T_{23}^2-D(2T_{14}-D)}{2(T_{14}-T_{23})D}\big)  & \text{if } T_{23} < D \leq T_{14},\\
1 - \delta \big(\frac{T_{14}+T_{23}}{2D}\big) & \text{if } D > T_{14},
\end{cases}
\end{equation}

The mean out-of-transit intensity can be calculated in a similar way, but it's
clear that the best case scenario will occur when it's simply taken to equal
unity, thus maximising the $\Delta F$. Similarly, the optimal choice of
$D$ must lie in the range $[T_{23},T_{14}]$. Following the same
steps as with $\SNR_{\Delta F,\mathrm{BB}}$, this ideal case leads to

\begin{align}
\lim_{\Tout \to \infty} \SNR_{\Delta F,\mathrm{TB}} &= \Bigg(\frac{D (2 T_{14}-D) - T_{23}^2}{2 D (T_{14}-T_{23})}\Bigg) \Bigg(\frac{\delta}{\sigma_0}\Bigg) \sqrt{D}.
\label{eqn:SNRDeltaFTB}
\end{align}

If we evaluate Equation~(\ref{eqn:SNRDeltaFTB}) in the case of $D \to T_{14}$,
we recover the SNR equation for a transit presented by \citet{ks:2016}
in their Equation~(11):

\begin{align}
\lim_{D \to T_{14}} \lim_{\Tout \to \infty} \SNR_{\Delta F,\mathrm{TB}} &= \Bigg(\frac{T_{14}+T_{23}}{2 \sqrt{T_{14}}}\Bigg) \Bigg(\frac{\delta}{\sigma_0}\Bigg).
\label{eqn:SNRDeltaFTB_KS}
\end{align}

This reproduction of the \citet{ks:2016} formula more clearly elucidates the
derivation and thinking behind it. The assumption is that the generative model
is a trapezoid, but the SNR is defined using a box-car delta-flux approach
(hence a TB scenario) where $D=T_{14}$. However, one can actually increase the
SNR here slightly by instead of setting $D=T_{14}$, optimising
$\SNR_{\Delta F,\mathrm{TB}}$ with respect to $D$ through differentiation,
yielding an optimal $D$ of

\begin{align}
\hat{D}_{\mathrm{\Delta F}} &= \frac{T_{14} + \sqrt{T_{14}^2 + 3 T_{23}^2}}{3},
\end{align}

where the hat notation denotes an optimised value, and the subscript $\Delta F$
is added to emphasise that this is only demonstrated to be the optimal choice
of $D$ when using the $\Delta F$ method.

This result may be useful for those running BLS in archival data on signals
of known $T_{23}$ and $T_{14}$, where the optimal signal width should be
set to the above to maximize SNR \citep{yao:2019}. At the optimal $D$, and in
the limit of large $\Tout$, one obtains

\begin{align}
\lim_{D \to \hat{D}_{\mathrm{\Delta F}}} \lim_{\Tout \to \infty} \SNR_{\Delta F,\mathrm{TB}} =& 
\Bigg(\frac{2\big(-3T_{23}^2+T_{14}(T_{14} + \sqrt{T_{14}^2+3T_{23}^2}) \big) }{3\sqrt{3}(T_{14}-T_{23}) \sqrt{T_{14}^2+3T_{23}^2}}\Bigg) \nonumber\\
\qquad& \times \Bigg(\frac{\delta}{\sigma_0}\Bigg).
\label{eqn:SNRDeltaFTB_hat}
\end{align}

It was verified that the same result is obtained if one does not assume
$\Iout=1$ but actually works through the cases, and of course the equivalence
is not surprising since in the limit of $\Tout \to \infty$ one necessarily has
$\Iout \to 1$.

\subsection{The Trapezoidal-Trapezoidal (TT) Approximation}

Finally, let us consider the case where the generative model is a trapezoid
and the model of comparison is also a trapezoid. The SNR cannot be computed
using the $\Delta F$ method here, so instead we use the $\Delta\chi^2$
approach. Feeding Equation~(\ref{eqn:ftrap}) into Equation~(\ref{eqn:SNRchi2}),
we obtain

\begin{align}
\mathrm{SNR}_{\Delta\chi^2,\mathrm{TT}} &= \sqrt{ \frac{T_{14}+2T_{23}}{3} } \Bigg( \frac{\delta}{\sigma_0} \Bigg).
\label{eqn:SNRfinal}
\end{align}

This is actually a far easier calculation than TB, which is greatly complicated
by the fact one is attempting to compare the wrong model to a generative
function. Our expectation here should be that using the correct model will lead
to a higher SNR, since otherwise one has a misspecified likelihood function and
that will pull the SNR down. Although limb darkening and ingress/egress
curvature are ignored here, largely because they are intractable to the
analytic approach sought in this paper, the example of limb darkening
exemplifies this via the Transit Least Squares (TLS) algorithm which achieves
slightly improved sensitivity over BLS by merely employing an improved transit
shape \citep{hippke:2019}.

\begin{figure}
\begin{center}
\includegraphics[width=8.6cm,angle=0,clip=true]{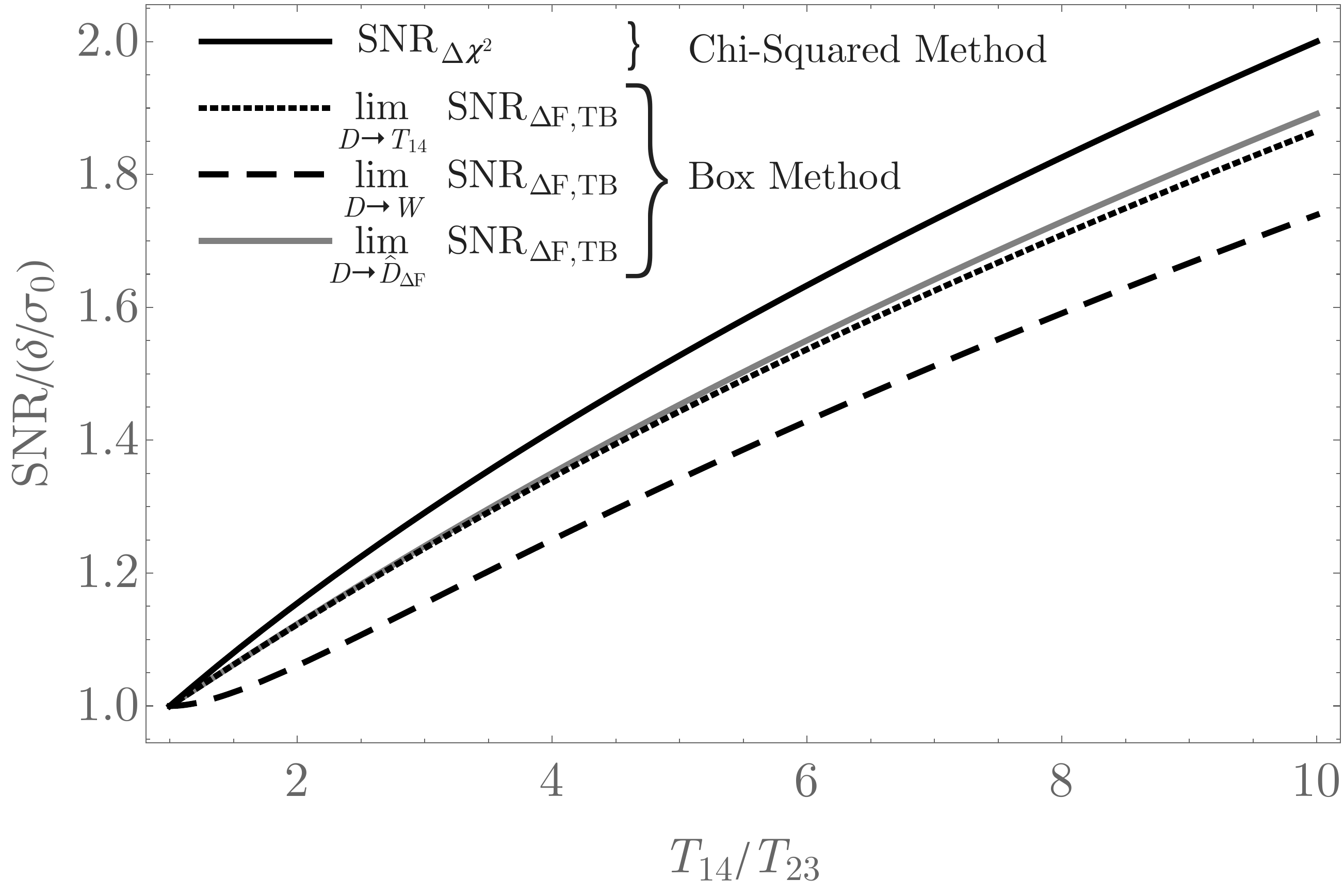}
\caption{\emph{
Comparison of four formulae for the SNR of a trapezoidal transit. The
$\Delta\chi^2$ method is the only one that uses the correct model to
describe the light curve shape, whereas $\Delta F$ (a box-car model)
is always an approximation to the trapezoid only. As a result, it always
produces worse SNRs than the $\Delta \chi^2$ method.}}
\label{fig:trapcomp}
\end{center}
\end{figure}

Figure~\ref{fig:trapcomp} verifies our expectations, where one can see four
versions of the SNR formulae applied to a trapezoid, three from the $\Delta F$
(TB) methods, and one from the $\Delta\chi^2$ (TT) method. Note how a BB
formula is not shown here (box-car models applied to box-car generated
transits) since this is conditioned upon a different data set (or really
generative model) to begin with. The formulae all converge when $T_{14} \to
T_{23}$, since in this limit the generative model becomes a box-car, and
as shown earlier in Section~\ref{sub:BB} one finds perfect agreement between
$\Delta F$ and $\Delta \chi^2$ methods here (since both are correctly
specified models essentially). In all other cases though, the $\Delta\chi^2$
method yields the highest SNR. In other words, any other formulae is in
fact underestimating the true SNR.

\subsection{A Note on Grazing Transits}

Having established that the $\Delta\chi^2$ method maximises SNR, let us now
consider how to extend to grazing transit geometries. In this case,
the impact parameter, $b$, lies in the range $(1-p)\leq b<(1+p)$, where $p \equiv
(R_P/R_{\star})$ is the ratio-of-radii. Accordingly, $T_{23}\to0$ but $T_{14}$
remains positive. For a non-grazing geometry (and ignoring limb darkening
as has been done throughout), $\delta=p^2$, but this breaks down for grazing
cases, becoming \citep{mandel:2002}

\begin{equation}
\delta[p,b] =
\begin{cases}
p^2  & \text{if } 0 \leq b < 1-p,\\
\frac{1}{\pi}\Big( p^2\kappa_0 + \kappa_1 - \sqrt{\frac{4b^2-(1+b^2-p^2)^2}{4}} \Big)  & \text{if } 1-p \leq b < 1+p,\\
0  & \text{if } 1+p \leq b < \infty,
\end{cases}
\label{eqn:fbox2}
\end{equation}

where

\begin{align}
\kappa_0 &= \cos^{-1}\Big(\frac{p^2+b^2-1}{2pb}\Big),\\
\kappa_1 &= \cos^{-1}\Big(\frac{1-p^2+b^2}{2b}\Big).
\end{align}

Thus, Equation~(\ref{eqn:SNRfinal}) remains valid except that it is understood
that $\delta$ is given by Equation~(\ref{eqn:fbox2}) and $T_{23}\to0$.

\subsection{A Note On Multiple Transits}

In the case of multiple transits, the analytic formulae presented thus far and
throughout are easily modified such that $\sigma_0 \to \sigma_0/\sqrt{N}$, where
$N$ is the number of transits observed. As before, this assumes regular sampled,
homoscedastic data. Real-world cases affected by partial transits,
irregular/sparsely sampling and/or heteroscedastic data may need to instead
numerically evaluate the SNR, as discussed earlier.

\section{The Sinning of Binning}
\label{sec:binning}

\subsection{The Effects of Binning}

Photometric time series are not captured with exposures of infinitesimal
duration. Each data point is an exposure of finite duration and during that
exposure the light curve can change. What this means is that our observations
are in fact average intensities over the exposure times. If one phase-folds a
large number of transits together with a certain exposure time, the resulting
light curve will be distorted from the naive model one might expect ignoring
this effect. Like a taking a photo of a racing car, the image appears blurred,
smeared out; and so too is a transit. A transit is particularly susceptible to
this because of its sharp discontinuities; observations captured near the first
and fourth contact points for example will not exhibit the crisp discontinuity
but rather a smoothed curve as a result of exposure time. It's important to
emphasise that all observations are in fact binned like this; one cannot ever
observe a transit with infinitesimal exposures.

This effect was first pointed out, in the case of transits, in
\citet{binning:2010} with the monicker ``binning is sinning''. And since then
the effect has been demonstrated and explored for supernovae \citep{brout:2021}
and phase curves \citep{morello:2022}. In \citet{binning:2010}, the effect was
explored numerically. Trapezoidal transits were simulated, smoothed with a
moving average filter, and then the consequences for parameter retrieval
were discussed, as well presenting numerical schemes to compensate. A basic
observation is that the observed first-to-fourth contact duration satisfies
$T_{14,\mathrm{obs}} = T_{14} + I$ and the observed second-to-third
contact duration satisfies $T_{23,\mathrm{obs}} = T_{23} - I$, where
$I$ is the exposure/integration time. But what is the SNR of this smeared out
transit?

Since the contact points shift in/out by $I$, one might be tempted to use
Equation~(\ref{eqn:SNRfinal}) but simply apply the appropriate modification to
$T_{14}$ and $T_{23}$. But this is wrong! Inspection of Figure~1 of
\citet{binning:2010} reveals why - it's not just that the contact
points shift, the transit shape has changed from a trapezoid (characterised
by straight lines) to a curved form i.e. $\mathrm{d}^2f/\mathrm{d}t\neq0$. In
other words, the generative model is no longer a trapezoid. So using a
trapezoidal model to calculate the SNR of a non-trapezoidal shape is wrong, as
was demonstrated in the last section.

What is needed is a closed-form expression for the shape of a trapezoidal
transit in the presence of finite integration time. \citet{binning:2010}
never presented such a formula, largely because it's cumbersome to
calculate and was unnecessary to the point of that paper. But here, we
cannot calculate the SNR of such a transit without it - at least not
analytically.

\subsection{The Convolved Transit}

The solution to this is to consider what a finite exposure time does.
Upon a phase-folded time series of transit observations, the finite
exposure time effect acts as a moving average with a bandwidth of
$I$. A moving average of some function, $f[t]$, is the convolution of
$f[t]$ with a top-hat function, $g[x]$:

\begin{align}
f'[t] &= f[t] \star g[t],\nonumber\\
\qquad &= \int_{-\infty}^{\infty} f[u] g[t-u]\,\mathrm{d}u.
\label{eqn:convolution}
\end{align}

This simple formula betrays the challenge of actually calculating this
for a trapezoid, because the trapezoid is a piece-wise function and
thus one has to essentially perform this convolution in a piece-meal
manner. In fact, there are nine unique regimes, which are best
illustrated in the case of a short exposure time but the existence of
nine regimes persists irrespective of $I$. If the
mid-exposure/integration time is given by $\ti$, then these are

\begin{enumerate}[i)]
\item $-\infty < \ti \leq t_1 - I/2$ (pre-transit),
\item $t_1 - I/2 < \ti \leq t_1 + I/2$ (saddles $1^{\mathrm{st}}$-contact),
\item $t_1 + I/2 < \ti \leq t_2 - I/2$ (within ingress),
\item $t_2 - I/2 < \ti \leq t_2 + I/2$ (saddles $2^{\mathrm{nd}}$-contact),
\item $t_2 + I/2 < \ti \leq t_3 - I/2$ (within totality),
\item $t_3 - I/2 < \ti \leq t_3 + I/2$ (saddles $3^{\mathrm{rd}}$-contact),
\item $t_3 + I/2 < \ti \leq t_4 - I/2$ (within egress),
\item $t_4 - I/2 < \ti \leq t_4 + I/2$ (saddles $4^{\mathrm{th}}$-contact),
\item $t_4 + I/2 < \ti \leq \infty$ (post-transit).
\end{enumerate}

One can now see what we mean by ``small'' $I$; it really defines the
integration time being less than any relevant timescale here, so smaller
than $T_{23}$ (which is necessarily smaller than $T_{14}$) and also smaller
than the ingress/egress time. Thus, the ``small'' $I$ regime is characterised
by $0<I<T_{23}<T_{14}$ and $I<(T_{14}-T_{23})/2$. In this regime, one can now
proceed through each of the nine cases and compute the new convolved transit,
$f_{\mathrm{T}}'[t]$, using Equation~(\ref{eqn:ftrap}) and

\begin{align}
f_{\mathrm{T}}'[t] &= I^{-1} \int_{\ti - I/2}^{\ti+I/2} f_{\mathrm{T}}[t]\,\mathrm{d}t,
\end{align}

which simplifies Equation~(\ref{eqn:convolution}) to the case of a top-hat
$g[t]$, as we have here. Besides from the small $I$ regime, this work has
identified five other regimes that each require a tailored treatment of
both the nine cases and the resulting integrals. Including the small $I$
regime, these are

\begin{enumerate}[A)]
\item $0<I<T_{23}<W<T_{14}$ and $0<I<(T_{14}-T_{23})/2$ (small $I$)
\item $0<I<T_{23}<W<T_{14}$ and $0<(T_{14}-T_{23})/2<I$ (ingress enveloping $I$)
\item $0<T_{23}<I<W<T_{14}$ and $0<I<(T_{14}-T_{23})/2$ (totality enveloping $I$)
\item $0<T_{23}<I<W<T_{14}$ and $0<(T_{14}-T_{23})/2<I$ (ingress-totality enveloping $I$)
\item $0<T_{23}<W<I<T_{14}$ [and $0<(T_{14}-T_{23})/2<I$] (FWHM enveloping $I$)
\item $0<T_{23}<W<T_{14}<I$ [and $0<(T_{14}-T_{23})/2<I$] (transit enveloping $I$)
\end{enumerate}

For the last two regimes, the second condition is put in square parentheses to
highlight that it's always satisfied given the first condition. One may now
proceed through the $9 \times 6$ integrals by hand and evaluate each function,
then stitch the results together into a piece-wise function. The final
expression was graphically tested with live manipulate functions against
numerically smoothed light curves. The results are presented in the Appendix
for the sake of conciseness here. It is noted that cases 1) and 2) also appear
in Equation~(7) of \citet{price:2014}, but they do not consider the longer
$I$ cases necessary to fully generalise the convolved transit.

\subsection{When it comes to SNR, is binning sinning?}

Equipped with our convolved transit expression, we can now finally come back to
the topic at the core of this paper - what is the SNR? This is crucial question
when it comes to the design of upcoming transit surveys, such as PLATO
\citep{rauer:2014}. The advantage of binning time series into ever coarser
integrations is that less data needs to be stored and transmitted back to Earth
from space-based missions, thus allowing a larger number of stars to be
observed. The disadvantage of larger integration times is that one smears out
the transit shape and thus have fundamentally lost information.

This loss of information has been demonstrated to cause a precision loss in the
retrieved transit parameters \citep{price:2014}. On this basis, it stands to
reason that the SNR will also be deteriorated due to the integration time effect.
Using our expressions for the trapezoid light curve with smearing, one may
evaluate the SNR rigorously using Equation~(\ref{eqn:SNRchi2}), which yields
the following piece-wise results:

\begin{align}
\Bigg(\frac{\SNR_{\Delta\chi^2,A}'}{\delta/\sigma_0}\Bigg)^2 =& \Big( 5 T_{23} I^2-5 T_{14} \left(3 T_{23}^2+I^2\right)+5 T_{14}^3+10 T_{23}^3\nonumber\\
\qquad&+2 I^3\Big)/\Big(15(T_{14}-T_{23})^2\Big)
\end{align}
\begin{align}
\Bigg(\frac{\SNR_{\Delta\chi^2,B}'}{\delta/\sigma_0}\Bigg)^2 =& \Big(
120 T_{23} I^2+T_{14} \left(20 T_{23} I+3 T_{23}^2+120 I^2\right)\nonumber\\
&-10 T_{23}^2 I-T_{14}^2\left(3 T_{23}+10 I\right)+T_{14}^3-T_{23}^3\nonumber\\
\qquad&-80 I^3 \Big)/\Big(240 I^2 \Big)
\end{align}
\begin{align}
\Bigg(\frac{\SNR_{\Delta\chi^2,C}'}{\delta/\sigma_0}\Bigg)^2 =& \Big(
-5 T_{14} I^4+5 T_{14}^3 I^2-5 T_{23}^2 I^2 \left(3 T_{14}-2 I\right)\nonumber\\
\qquad&+5 T_{23}^4 I-T_{23}^5+3 I^5 \Big)/\Big( 15 (T_{14}-T_{23})^2 I^2 \Big)
\end{align}
\begin{align}
\Bigg(\frac{\SNR_{\Delta\chi^2,D}'}{\delta/\sigma_0}\Bigg)^2 =& \Big(
-80 T_{23} I^4+80 T_{23}^2 I^3-40 T_{23}^3 I^2\nonumber\\
\qquad&+10 T_{14}^3 \left(4 T_{23} I+T_{23}^2+12 I^2\right)\nonumber\\
\qquad&+5 T_{14} T_{23} \left(-24 T_{23} I^2+8 T_{23}^2 I+T_{23}^3+32 I^3\right)\nonumber\\
\qquad&+70 T_{23}^4 I-10 T_{14}^2 \left(T_{23}+2 I\right){}^3\nonumber\\
\qquad&-5 T_{14}^4 \left(T_{23}+2 I\right)+T_{14}^5-17T_{23}^5\nonumber\\
\qquad&+16 I^5 \Big)/\Big( 240 (T_{14}-T_{23})^2 I^2 \Big)
\end{align}
\begin{align}
\Bigg(\frac{\SNR_{\Delta\chi^2,E}'}{\delta/\sigma_0}\Bigg)^2 =& \Big(
10 T_{14}^3 \left(T_{23}^2+8 I^2\right)-20 T_{14}^2 \left(3 T_{23}^2 I+4 I^3\right)\nonumber\\
\qquad&+5 T_{14} \left(T_{23}^4+8 I^4\right)-10 T_{14}^4 I+30 T_{23}^4 I\nonumber\\
\qquad& +T_{14}^5-8 T_{23}^5-8 I^5 \Big)/\Big( 120 (T_{14}-T_{23})^2 I^2 \Big)
\end{align}
\begin{align}
\Bigg(\frac{\SNR_{\Delta\chi^2,F}'}{\delta/\sigma_0}\Bigg)^2 =& \Big(
T_{14}^2 \left(30 I-14 T_{23}\right)+T_{23} T_{14} \left(60 I-11 T_{23}\right)\nonumber\\
\qquad&+30 T_{23}^2 I-7 T_{14}^3-8T_{23}^3 \Big)/\Big( 120 I^2 \Big)
\end{align}

\begin{figure}
\begin{center}
\includegraphics[width=8.6cm,angle=0,clip=true]{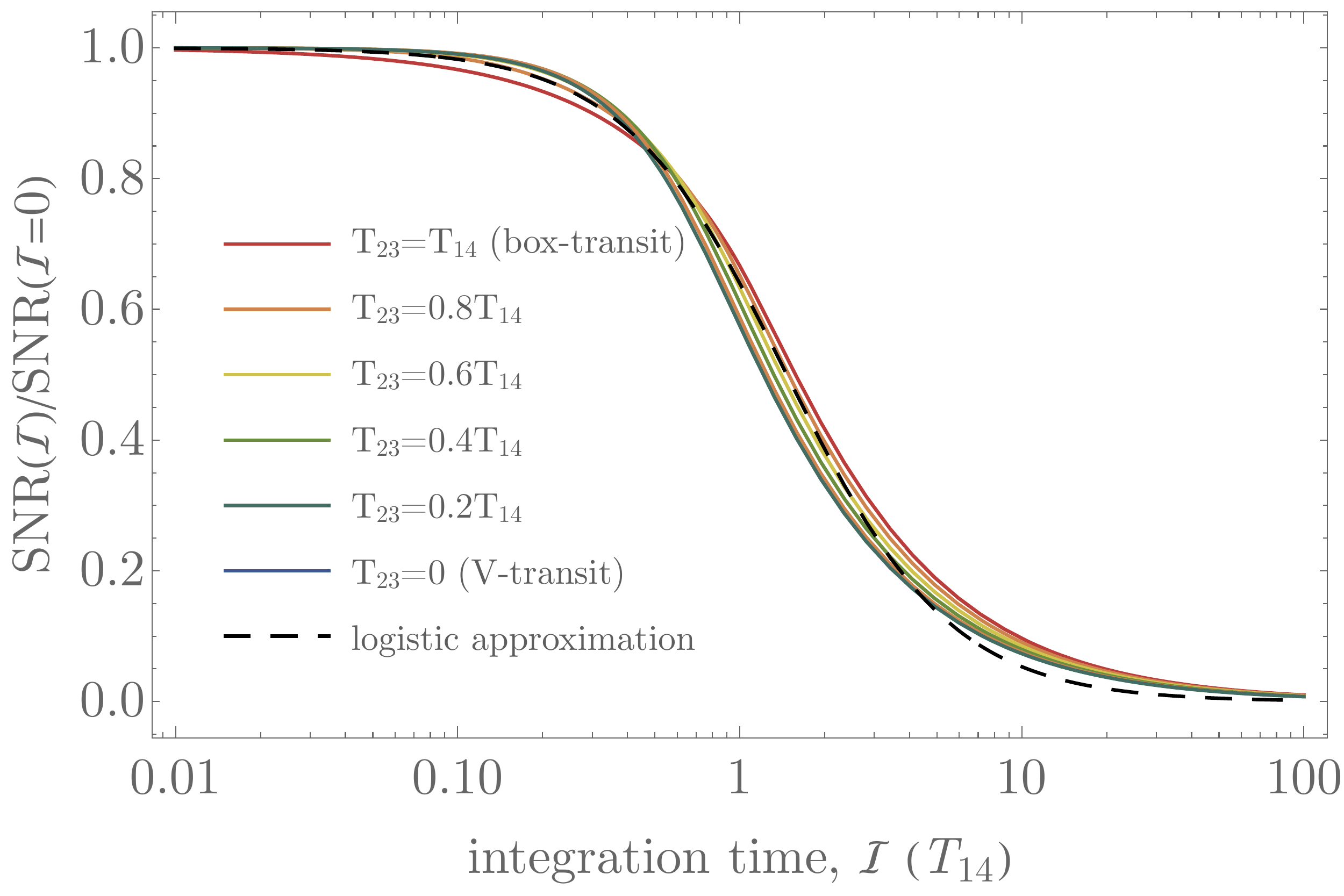}
\caption{\emph{
SNR of a trapezoidal transit with finite integration time, relative to
the SNR found with infintessimal integration time. As expected, integration
time effects attenuate the transit SNR and can be well approximated by
a logistic function given by Equation~(\ref{eqn:logistic}).
}}
\label{fig:SNRsmear}
\end{center}
\end{figure}

Rather than trying to gain insight from the equations themselves, it's far more
intuitive to simply plot the smeared SNR as a function of integration time, as
shown in Figure~\ref{fig:SNRsmear}. Note that this figure normalises the
smeared SNR by the original expression found earlier ignoring this effect
(Equation~\ref{eqn:SNRfinal}), since the depth and measurement uncertainty
consistently cancel out in doing so. Thus, the loss in SNR is completely
described by $T_{14}$, $T_{23}$ and $I$. By working in units of $T_{14}$, one
can further reduce the dimensionality to two. This is visualised by using the
$x$-axis for one of these ($I/T_{14}$) and multiple colored lines for the other
($T_{23}/T_{14}$).

As expected, the SNR drops off smoothly and monotonically in all cases,
providing some assurance that our piece-wise function correctly connects at the
case boundaries, as well exhibiting the intuitive behaviour one expects of SNR
dropping with ever larger $I$. Crucially, the SNR never drops to zero even at
excessive large $I$ values, although it does asymptotically approach zero (again
as expected). It is also worth noting that the ensemble of models are in fairly
good agreement with one another, and can be approximated by a single function
shown in black, and given by

\begin{align}
\frac{\SNR(I)}{\SNR(I=0)} \simeq \frac{1}{1+\exp\big(-\tfrac{3}{2}(0.384363-\log_e\tfrac{I}{T_{14}})\big)}.
\label{eqn:logistic}
\end{align}

The approximation dictates that one loses $\gtrsim10$\% of the SNR when
$I \gtrsim T_{14}/3$, which serves as a useful rule of thumb
to guide observers. In extreme cases, such as 30-minute \textit{Kepler}
long-cadence on a 1-minute white dwarf transit, the SNR loss is severe and
drops to 1\% of the unsmeared value.

\subsection{A Note On Extreme Integration Times}

Recall earlier in Section~\ref{sub:SNRcontinuous} how an underlying assumption
of our analytic formulae is that the transit is well-sampled, which comes into
question as the integration time increases. For sparsely sampled transits, the
phasing of the points with respect to the transit mid-time will increasingly
influence the resulting SNR. However, even when $\mathcal{I} \sim W$, the
problem of special phasing arrangements washes out if $N\gg1$, since the
phase-folded transit will again have nearly uniform, regular sampling. In this
sense, a single transit with sparse sampling represents the most egregious
violation of our sampling assumption.

\section{But Transits Aren't Trapezoids!}
\label{sec:discussion}

This work has discussed how, within the existing exoplanet literature,
the most commonly used formula for transit SNR is only accurate for box-shaped
transits. Further, a transit can never be (naturally) box-shaped
\citep{cloaking:2016}. The box-shaped transit can perhaps be best described as
a zeroth-order approximation of some physically motivated light curve (e.g.
\citealt{mandel:2002}), and in that vein a trapezoidal light curve is a refined
approximation - but unquestionably still an approximation. In this sense, the
new SNR formula presented in this work for a trapezoidal light curve is also
an approximation (Equation~\ref{eqn:SNRfinal}), albeit an improved
approximation upon the box.

So what if one wishes to go further and have a light curve model that correctly
accounts for the ingress/egress area of overlap and/or limb darkening effects
\citep{mandel:2002}? Or indeed, what if one wishes to use an even more
sophisticated model than that; one which accounts for gravity darkening
\citep{herman:2018}, planetary oblateness \citep{carter:2010}, nightside
pollution \citep{nightside:2010}, atmospheric refraction \citep{sidis:2010} and
indeed many other possible complications? In this work, the author was not
able to find a closed-form SNR solution (using the integral of
Equation~\ref{eqn:SNRchi2}) to even the relatively simple case of a non-limb
darkened \citet{mandel:2002} model and thus a trapezoidal transit may be
the most accurate approximate model one can ever hope to have a closed form
SNR transit equation for.

A closed-form solution like this certainly has its place though, and for
surveys operating towards the infrared (where limb darkening is suppressed)
it may be wholly appropriate. But in cases where more finesse is required, it
is recommended to numerically evaluate the SNR of the transit model using
Equation~(\ref{eqn:SNRchi2}) for continuous functions in the presence of
regularly sampled, homoscedastic data, or Equation~(\ref{eqn:discrete})
otherwise.

More broadly, this work establishes that the boxcar transit SNR is
not magically equivalent to the trapezoidal case, nor indeed more nuanced
transit models. So adopting the boxcar transit SNR comes at some finite
loss in accuracy. Section~\ref{sec:binning} also establishes that using
longer exposure/binning/integration times in a photometric survey will
reduce the SNR of transits. The crucial timescale here is $T_{14}$.
For integrations times much faster than this, the reduction is small,
whereas for times longer than this, there is a dramatic decrease.
This is particularly pertinent when considering the detection efficiency
towards very short transients, such as white dwarf transiting planets
\citep{vanderburg:2020}, whose transits typically last for mere minutes
\citep{cortes:2019}. Even the ``fast'' 2-minute cadence of \textit{TESS}
\citep{ricker:2015} is too slow to avoid substantial SNR loss  here.
Any effort at occurrence rate calculations for such transients (e.g.
\citealt{sluijs:2018}) must be careful to not only use a realistic
astrophysical model, but also account for the convolutional smearing,
which will surely require a numerical approach for full accuracy.
	
\section*{Acknowledgments}

Thank-you to the anonymous reviewer for their helpful review.
This work was supported by patreons to the Cool Worlds Lab, for which the author thanks
D. Smith, M. Sloan, C. Bottaccini, D. Daughaday, A. Jones, S. Brownlee, N. Kildal, Z. Star, E. West, T. Zajonc, C. Wolfred, L. Skov, G. Benson, A. De Vaal, M. Elliott, B. Daniluk, M. Forbes, S. Vystoropskyi, S. Lee, Z. Danielson, C. Fitzgerald, C. Souter, M. Gillette, T. Jeffcoat, J. Rockett, D. Murphree, S. Hannum, T. Donkin, K. Myers, A. Schoen, K. Dabrowski, J. Black, R. Ramezankhani, J. Armstrong, K. Weber, S. Marks, L. Robinson, S. Roulier, B. Smith, G. Canterbury, J. Cassese, J. Kruger, S. Way, P. Finch, S. Applegate, L. Watson, E. Zahnle, N. Gebben, J. Bergman, E. Dessoi, J. Alexander, C. Macdonald, M. Hedlund, P. Kaup, C. Hays, W. Evans, D. Bansal, J. Curtin, J. Sturm, RAND Corp., M. Donovan, N. Corwin, M. Mangione, K. Howard, L. Deacon, G. Metts, G. Genova, R. Provost, B. Sigurjonsson, G. Fullwood, B. Walford, J. Boyd, N. De Haan, J. Gillmer, R. Williams, E. Garland, A. Leishman, A. Phan Le, R. Lovely, M. Spoto, A. Steele, M. Varenka, K. Yarbrough, A. Cornejo, D. Compos, F. Demopoulos, G. Bylinsky, J. Werner, B. Pearson, S. Thayer, T. Edris \& M. Waters.

\section*{Data Availability}

This work did not produce or use any data.

\onecolumn
\appendix
\section*{Appendix}

This appendix provides the equations describing the light curve shape of a
trapezoidal transit convolved with a top-hat i.e. a trapezoidal light curve
observed with finite integration time, $g_{\mathrm{T}}[t]$.

\begin{equation}
g_{\mathrm{T}}[t] =
\begin{cases}
g_{\mathrm{T},\mathrm{A}}[t]
& \text{if } 0<I<T_{23}<W<T_{14} \text{ and } 0<I<(T_{14}-T_{23})/2,\\
g_{\mathrm{T},\mathrm{B}}[t]
& \text{if } 0<I<T_{23}<W<T_{14} \text{ and } 0<(T_{14}-T_{23})/2<I,\\
g_{\mathrm{T},\mathrm{C}}[t]
& \text{if } 0<T_{23}<I<W<T_{14} \text{ and } 0<I<(T_{14}-T_{23})/2,\\
g_{\mathrm{T},\mathrm{D}}[t]
& \text{if } 0<T_{23}<I<W<T_{14} \text{ and } 0<(T_{14}-T_{23})/2<I,\\
g_{\mathrm{T},\mathrm{E}}[t]
& \text{if } 0<T_{23}<W<I<T_{14} \text{ and } 0<(T_{14}-T_{23})/2<I,\\
g_{\mathrm{T},\mathrm{F}}[t]
& \text{if } 0<T_{23}<W<T_{14}<I \text{ and } 0<(T_{14}-T_{23})/2<I,
\end{cases}
\label{eqn:gmaster}
\end{equation}

\begin{equation}
g_{\mathrm{T},\mathrm{A}}[t] =
\begin{cases}
1
& \text{if } t \leq (-T_{14}-I)/2,\\
1-\frac{\delta (2 t+T_{14}+I)^2}{4 I (T_{14}-T_{23})}
& \text{if }  (-T_{14}-I)/2 <t \leq (I-T_{14})/2,\\
1-\frac{\delta (2 t+T_{14})}{T_{14}-T_{23}}
& \text{if } (I-T_{14})/2 < t\leq  (-T_{23}-I)/2,\\
\frac{\delta  \left(4 t^2+4 t (T_{23}-I)+(T_{23}+I)^2\right)-4 (\delta -1) T_{14} I-4T_{23} I}{4 I (T_{14}-T_{23})}
& \text{if } (-T_{23}-I)/2 < t \leq (I-T_{23})/2,\\
1-\delta
& \text{if } (I-T_{23})/2 < t \leq (T_{23}-I)/2,\\
\frac{\delta \left(4 t^2+4 t (I-T_{23})+(T_{23}+I)^2\right)-4 (\delta -1) T_{14} I-4 T_{23} I}{4 I (T_{14}-T_{23})}
& \text{if } (T_{23}-I)/2 < t \leq (T_{23}+I)/2,\\
\frac{\delta  (2 t-T_{14})}{T_{14}-T_{23}}+1
& \text{if} (T_{23}+I)/2 < t \leq (T_{14}-I)/2,\\
1-\frac{\delta  (-2 t+T_{14}+I)^2}{4 I (T_{14}-T_{23})}
& \text{if} (T_{14}-I)/2 < t \leq (T_{14}+I)/2 \\
1
& \text{if } t>\frac{T_{14}+I}{2}.
\end{cases}
\label{eqn:ga}
\end{equation}

\begin{equation}
g_{\mathrm{T},\mathrm{B}}[t] =
\begin{cases}
1
& \text{if } t \leq (-T_{14}-I)/2,\\
1-\frac{\delta (2 t+T_{14}+I)^2}{4 I (T_{14}-T_{23})}
& \text{if }  (-T_{14}-I)/2 < t \leq (-T_{23}-I)/2,\\
1-\frac{\delta (4 t+T_{14}+T_{23}+2 I)}{4 I}
& \text{if } (-T_{23}-I)/2 t \leq (I-T_{14})/2,\\
\frac{\delta \left(4 t^2+4 t (T_{23}-I)+(T_{23}+I)^2\right)-4 (\delta -1) T_{14} I-4T_{23} I}{4 I (T_{14}-T_{23})}
& \text{if } (I-T_{14})/2 < t \leq (I-T_{23})/2,\\
1-\delta
& \text{if } (I-T_{23})/2 < t \leq (T_{23}-I)/2,\\
\frac{\delta \left(4 t^2+4 t (I-T_{23})+(T_{23}+I)^2\right)-4 (\delta -1) T_{14} I-4T_{23} I}{4 I (T_{14}-T_{23})}
& \text{if } (T_{23}-I)/2 < t \leq (T_{14}-I)/2,\\
1-\frac{\delta (-4 t+T_{14}+T_{23}+2 I)}{4 I}
& \text{if} (T_{14}-I)/2 < t \leq (T_{23}+I)/2,\\
1-\frac{\delta (-2 t+T_{14}+I)^2}{4 I (T_{14}-T_{23})}
& \text{if} (T_{23}+I)/2 < t \leq (T_{14}+I)/2,\\
1
& \text{if } t>(T_{14}+I)/2.
\end{cases}
\label{eqn:gb}
\end{equation}

\begin{equation}
g_{\mathrm{T},\mathrm{C}}[t] =
\begin{cases}
1
& \text{if } t \leq (-T_{14}-I)/2,\\
1-\frac{\delta (2 t+T_{14}+I)^2}{4 I (T_{14}-T_{23})}
& \text{if } (-T_{14}-I)/2<t<(I-T_{14})/2,\\
1-\frac{\delta (2 t+T_{14})}{T_{14}-T_{23}}
& \text{if } (I-T_{14})/2 < t \leq (-T_{23}-I)/2,\\
\frac{\delta \left(4 t^2+4 t (T_{23}-I)+(T_{23}+I)^2\right)-4 (\delta -1) T_{14} I-4 T_{23} I}{4 I (T_{14}-T_{23})}
& \text{if } (-T_{23}-I)/2 < t \leq (T_{23}-I)/2,\\
\frac{\delta \left(4 t^2+I (I-2 T_{14})+T_{23}^2\right)}{2 I (T_{14}-T_{23})}+1
& \text{if } (T_{23}-I)/2 < t \leq (I-T_{23})/2,\\
\frac{\delta \left(4 t^2+4 t (I-T_{23})+(T_{23}+I)^2\right)-4 (\delta -1) T_{14} I-4 T_{23} I}{4 I (T_{14}-T_{23})}
& \text{if } (I-T_{23})/2 < t \leq (T_{23}+I)/2,\\
\frac{\delta (2 t-T_{14})}{T_{14}-T_{23}}+1
& \text{if} (T_{23}+I)/2 < t \leq (T_{14}-I)/2,\\
1-\frac{\delta  (-2 t+T_{14}+I)^2}{4 I (T_{14}-T_{23})}
& \text{if} (T_{14}-I)/2 < t \leq (T_{14}+I)/2,\\
1
& \text{if } t>(T_{14}+I)/2.
\end{cases}
\label{eqn:gc}
\end{equation}

\begin{equation}
g_{\mathrm{T},\mathrm{D}}[t] =
\begin{cases}
1
& \text{if } t \leq (-T_{14}-I)/2,\\
1-\frac{\delta  (2 t+T_{14}+I)^2}{4 I (T_{14}-T_{23})}
& \text{if } (-T_{14}-I)/2 < t \leq (-T_{23}-I)/2,\\
1-\frac{\delta (4 t+T_{14}+T_{23}+2 I)}{4 I}
& \text{if } (-T_{23}-I)/2 < t \leq (I-T_{14})/2,\\
\frac{\delta \left(4 t^2+4 t (T_{23}-I)+(T_{23}+I)^2\right)-4 (\delta -1) T_{14} I-4 T_{23} I}{4 I (T_{14}-T_{23})}
& \text{if } (I-T_{14})/2 < t \leq (T_{23}-I)/2,\\
\frac{\delta \left(4 t^2+I (I-2 T_{14})+T_{23}^2\right)}{2 I (T_{14}-T_{23})}+1
& \text{if } (T_{23}-I)/2 < t \leq (I-T_{23})/2,\\
\frac{\delta \left(4 t^2+4 t (I-T_{23})+(T_{23}+I)^2\right)-4 (\delta -1) T_{14} I-4T_{23} I}{4 I (T_{14}-T_{23})}
& \text{if } (I-T_{23})/2 < t \leq (T_{14}-I)/2,\\
1-\frac{\delta (-4 t+T_{14}+T_{23}+2 I)}{4 I}
& \text{if} (T_{14}-I)/2 < t \leq (T_{23}+I)/2,\\
1-\frac{\delta (-2 t+T_{14}+I)^2}{4 I (T_{14}-T_{23})}
& \text{if} (T_{23}+I)/2 < t \leq (T_{14}+I)/2,\\
1
& \text{if } t>(T_{14}+I)/2.
\end{cases}
\label{eqn:gd}
\end{equation}

\begin{equation}
g_{\mathrm{T},\mathrm{E}}[t] =
\begin{cases}
1
& \text{if } t \leq (-T_{14}-I)/2,\\
1-\frac{\delta (2 t+T_{14}+I)^2}{4 I (T_{14}-T_{23})}
& \text{if } (-T_{14}-I)/2 < t \leq (-T_{23}-I)/2,\\
1-\frac{\delta (4 t+T_{14}+T_{23}+2 I)}{4 I}
& \text{if } (-T_{23}-I)/2 < t \leq (T_{23}-I)/2,\\
1-\frac{\delta \left(2 T_{14} (2 t+I)-(2 t+I)^2+T_{14}^2-2 T_{23}^2\right)}{4 I
   (T_{14}-T_{23})}
& \text{if } (T_{23}-I)/2 < t \leq (I-T_{14})/2,\\
\frac{\delta \left(4 t^2+I (I-2 T_{14})+T_{23}^2\right)}{2 I (T_{14}-T_{23})}+1
& \text{if } (I-T_{14})/2 < t \leq (T_{14}-I)/2,\\
1-\frac{\delta \left(2 T_{14} (I-2 t)-(I-2 t)^2+T_{14}^2-2 T_{23}^2\right)}{4 I(T_{14}-T_{23})}
& \text{if } (T_{14}-I)/2 < t \leq (I-T_{23})/2,\\
1-\frac{\delta (-4 t+T_{14}+T_{23}+2 I)}{4 I}
& \text{if} (I-T_{23})/2 < t \leq (T_{23}+I)/2,\\
1-\frac{\delta (-2 t+T_{14}+I)^2}{4 I (T_{14}-T_{23})}
& \text{if} (T_{23}+I)/2 < t \leq (T_{14}+I)/2,\\
1
& \text{if } t>(T_{14}+I)/2.
\end{cases}
\label{eqn:ge}
\end{equation}

\begin{equation}
g_{\mathrm{T},\mathrm{F}}[t] =
\begin{cases}
1
& \text{if } t \leq (-T_{14}-I)/2,\\
1-\frac{\delta (2 t+T_{14}+I)^2}{4 I (T_{14}-T_{23})}
& \text{if } (-T_{14}-I)/2 < t \leq (-T_{23}-I)/2,\\
1-\frac{\delta (4 t+T_{14}+T_{23}+2 I)}{4 I}
& \text{if } (-T_{23}-I)/2 < t \leq (T_{23}-I)/2,\\
1-\frac{\delta \left(2 T_{14} (2 t+I)-(2 t+I)^2+T_{14}^2-2 T_{23}^2\right)}{4 I(T_{14}-T_{23})}
& \text{if } (T_{23}-I)/2 < t \leq (T_{14}-I)/2,\\
1-\frac{\delta (T_{14}+T_{23})}{2 I}
& \text{if } (T_{14}-I)/2 < t \leq (I-T_{14})/2,\\
1-\frac{\delta \left(2 T_{14} (I-2 t)-(I-2 t)^2+T_{14}^2-2 T_{23}^2\right)}{4 I
   (T_{14}-T_{23})}
& \text{if } (I-T_{14})/2 < t \leq (I-T_{23})/2,\\
1-\frac{\delta (-4 t+T_{14}+T_{23}+2 I)}{4 I}
& \text{if} (I-T_{23})/2 < t \leq (T_{23}+I)/2,\\
1-\frac{\delta  (-2 t+T_{14}+I)^2}{4 I (T_{14}-T_{23})}
& \text{if} (T_{23}+I)/2 < t \leq (T_{14}+I)/2,\\
1
& \text{if } t>(T_{14}+I)/2.
\end{cases}
\label{eqn:gf}
\end{equation}

\bsp
\label{lastpage}
\end{document}